\documentclass[pre,twocolumn]{revtex4}

\usepackage{amsmath}
\usepackage{amsfonts}
\usepackage{hyperref}
\usepackage{graphicx}

\begin{document}

\title{Dressing the Cusp: How Sharp-Edge Diffraction Theory Solves a Basic Issue in Catastrophe Optics}

\author{Riccardo Borghi}
\affiliation{Dipartimento di Ingegneria Civile, Informatica e delle Tecnologie Aeronautiche, \\
Universit\`{a} ``Roma Tre'', Via Vito Volterra 62, I-00146 Rome, Italy}

\begin{abstract}
The description of light diffraction using catastrophe optics is one of the most intriguing theoretical invention 
in the field of classical optics of the last four decades. Its practical implementation 
has faced some resistance over the years, 
mainly due to the difficulty of mathematically decorating the different, topologically speaking, types of optical singularities (caustics)
that concur to build up the skeleton on which diffraction patterns stem. 
Such a fundamental 
{\em dressing problem} has been solved in the past only for the so-called {\em fold},
which lies at the bottom of the hierarchy of structurally stable caustics. Climbing this hierarchy implies considerably more challenging mathematical problems to be solved.
\\
An ancient mathematical theorem is here employed to find the complete solution of the dressing problem
for the 
{\em cusp}, which is placed, in the stable caustic hierarchy, immediately after the fold. The other ingredient used for achieving such an important theoretical result is the paraxial version of the boundary diffraction wave theory, whose tight connection with
catastrophe optics has recently been emphasized in [R. Borghi, Opt. Lett. {\bf 41,} 3114 - 3117 (2016)].
\\
A significant example of the developed algorithm aimed at demonstrating its effectiveness and ease of implementation, is also presented.
\end{abstract}

\maketitle

\section{Introduction}
\label{Sec:Introduction}


It is an indisputable fact that light diffraction must be considered a wave phenomenon within a
classical (i.e., non-quantum) context. 
As far as the mathematical modeling of such a phenomenon is concerning, things are less certain.
Two main strategies can be identified for this purpose. The first strategy aims to build (and solve) suitable wave equations
starting from Maxwell's equations.
%
In 1817, Augustine Fresnel submitted a paper to the French Academy of Science, in which Huygens' principle was
mathematically implemented as a double integral, the celebrated Fresnel integral.
%
Fresnel's integral domain coincides with the entire wavefront, which acts as a secondary source within Huygens' scheme. When its domain is spatially limited by a planar sharp-edge aperture, Fresnel's integral can always be converted, under very general hypotheses, into a 1D contour integral defined on the aperture edge~\cite{Borghi/2022}. 
As a result, the computational complexity of the diffraction integrals is significantly reduced. 
But there is  more.

When a monochromatic plane wave impinges orthogonally on a planar, sharp-edge aperture of arbitrary shape, the above contour integral 
can always be written as the sum of two contributions, each of which has a sound physical interpretation
in the context of the so-called boundary diffraction wave (BDW) theory~\cite{Born/Wolf/1999,Borghi/2015}.
Surprisingly, the mathematical equivalence between the paraxial BDW theory and Fresnel's diffraction theory has been proved only recently, as witnessed by the author himself~\cite{Hannay/2000}:
\begin{quotation}
{\em ... it seems implausible that all this has escaped attention previously...}
\end{quotation}
%
The computational advantages are not the only reason for preferring paraxial BDW's theory over Fresnel.
In some cases, the former offered such a privileged conceptual viewpoint over the latter, that the description and interpretation of unexpected phenomena has been done in a way that could hardly be obtained in terms of Fresnel’s theory~\cite{Borghi/2017,Borghi/2018,Borghi/Carosella/2022}.

The attractiveness of paraxial BDW's theory lies in its natural ability to serve as a mathematical tool for studying light diffraction phenomena in a way that differs significantly from Huygens' principle.
About twenty years before the publication of Fresnel's memoir, Thomas Young had already proposed the revolutionary idea that the rim of an illuminated aperture could act as a secondary light source~\cite{Young/1802,Maggi/1888,Rubinowicz/1957}.    
Young's picture describes the diffraction of plane waves by sharp-edge apertures as a superposition of two fields.
The first field is produced by clipping the incoming wave according to the laws of geometrical optics.
The second field is originated by the aperture edge. These fields correspond to the contributions of BDW theory mentioned earlier.
In~\cite{Borghi/2015,Borghi/2016}, it was demonstrated that paraxial BDW theory provides a physically sound description of sharp-edge diffraction in the limit of non-small Fresnel numbers,
through the language of the so-called Catastrophe Optics (CO)~\cite{Berry/Upstill/1980,Nye/1999}.
CO is a powerful theoretical framework for modeling light behavior in real (i.e., nonideal) optical systems. Sharp-edge diffraction, for example, produces significant focusing effects, the probably most known being the Poisson (or Arago) spot,
which is generated on the axis of a perfectly circular opaque plate under plane-wave illumination.
In a sense, this is an ideal focusing system. However, the Poisson spot is highly unstable. In fact, it is sufficient to gently squeeze the circular plate to give it an elliptical shape of some eccentricity, say $\epsilon$, for the Poisson spot to inevitably ``explode'' into a four-cusp caustic. More importantly, such a configuration remains topologically {unchanged} for further perturbations of the aperture shape: always four cusps, for any $\epsilon\in (0,1)$.
%
What has been described is a clear example of {\em structural stability}, a topological concept introduced by Ren\'e Thom~\cite{Thom/1989}. Thom also proved a fundamental classification theorem within the context of catastrophe theory~\cite{Arnold/1986}. In optics, a structurally stable caustic is resistant to small perturbations of the external physical parameters that produce the diffracted field.  
After Thom's theorem was injected into wave optics, Catastrophe Optics was born~\cite{Berry/Upstill/1980,Nye/1999}.
It is our second strategy: rather than attempting to solve wave equations, CO's mathematical description of diffraction is
built up starting from scenarios made by optical singularities of several types,
like caustics and shadow boundaries. Each of them appears ``decorated,'' when observed at the wavelength scale, 
by a characteristic diffractive pattern. 
%
The following quote from John Nye captures the essence of CO~\cite{Nye/1999}:
\begin{quotation}
{\em Catastrophe optics is aimed at ``adding"  wave optics to geometrical optics in such a way that the wavefields so built are no longer divergent at caustics}
\end{quotation}
i.e.,~\cite{Berry/Upstill/1980},
\begin{quotation}
{\em at the most important places, where the light is brightest. }
\end{quotation}

Such caustic decoration is mathematically operated by suitable special functions, which are expressed in terms of certain canonical integrals, called \emph{diffraction catastrophes}~\cite{Berry/Upstill/1980},
\begin{equation}
     \label{canonical}
     \Psi(\boldsymbol{C})=\displaystyle\frac{1}{(2\pi)^{m/2}}\,
     \displaystyle\int\,\ldots\displaystyle\int_{\mathbb{R}^m}\,\mathrm{d}^{m}s
     \,\exp[\mathrm{i}\,\Phi(\boldsymbol{s};\boldsymbol{C})].
\end{equation}
Here $\Phi(\boldsymbol{s};\boldsymbol{C})$, the so-called generating function, is a {polynomial} (with degree $n\ge 3$) with respect to the integration variable $\boldsymbol{s}$ and {linear} with respect to the variable $\boldsymbol{C}$, the so-called {\em control state}.
The dimension of $\boldsymbol{C}$, is called \emph{codimension}. 

For the scope of the present paper, 
only the simplest types of catastrophes will be considered:  the \emph{fold}, denoted  $\mathcal{A}_2$ in the language of CO, 
and the \emph{cusp}, denoted by the symbol $\mathcal{A}_3$. Their generating functions $\Phi({s};\boldsymbol{C})$ in Eq.~(\ref{canonical}) are listed in Table~\ref{Tab:CausticClassification} and the corresponding diffraction catastrophes $\Psi$ turn out to be proportional to the Airy and to the Pearcey functions, respectively. 
\begin{table}[!ht]
\centering
\begin{tabular}{cc}
\hline
symbol s& $\Phi({s};\boldsymbol{C})$  \\
 \hline  \hline 
$\mathcal{A}_2$  &  ${s^3}\,+\,C_1\,s$ \\
$\mathcal{A}_3$  &  ${s^4}\,+\,C_2\,s^2\,+\,C_1\,s$ \\
\hline\hline
\end{tabular}
\caption{Generating functions for the fold ($\mathcal{A}_2$) and the cusp ($\mathcal{A}_3$) diffraction catastrophes considered in the present paper.}
\label{Tab:CausticClassification}
\end{table}

To complete the CO description of a diffracted wavefield near a given singularity, the following fundamental problem must then be solved:
\begin{quotation}
to express the control state $\boldsymbol{C}$ in terms of the {\em geometrical} parameters which mathematically  describe the singularity 
iteself
\end{quotation}
In~\cite{Borghi/2015,Borghi/2016,Borghi/2017}, the full analytical solution to this problem was found for the fold singularity $\mathcal{A}_2$. In the present paper, such result will be extended to the, considerably more challenging scenario of cusp singularities $\mathcal{A}_3$. To this end, use will be made by the so-called {\em Girard–Newton identities}, an important algebraic tool of group theory. These identities can be traced back to one of the several beautiful mathematical results achieved by Isaac Newton during his {\em annus mirabilis} 1666.  

Before proceeding, it is worth briefly summarizing paraxial BDW theory.
 
\section{A short {\em tour} on paraxial BDW theory}
\label{Sec:BDWTheory}

\begin{figure}[!ht]
\centerline{\includegraphics[width=4.5cm,angle=-90]{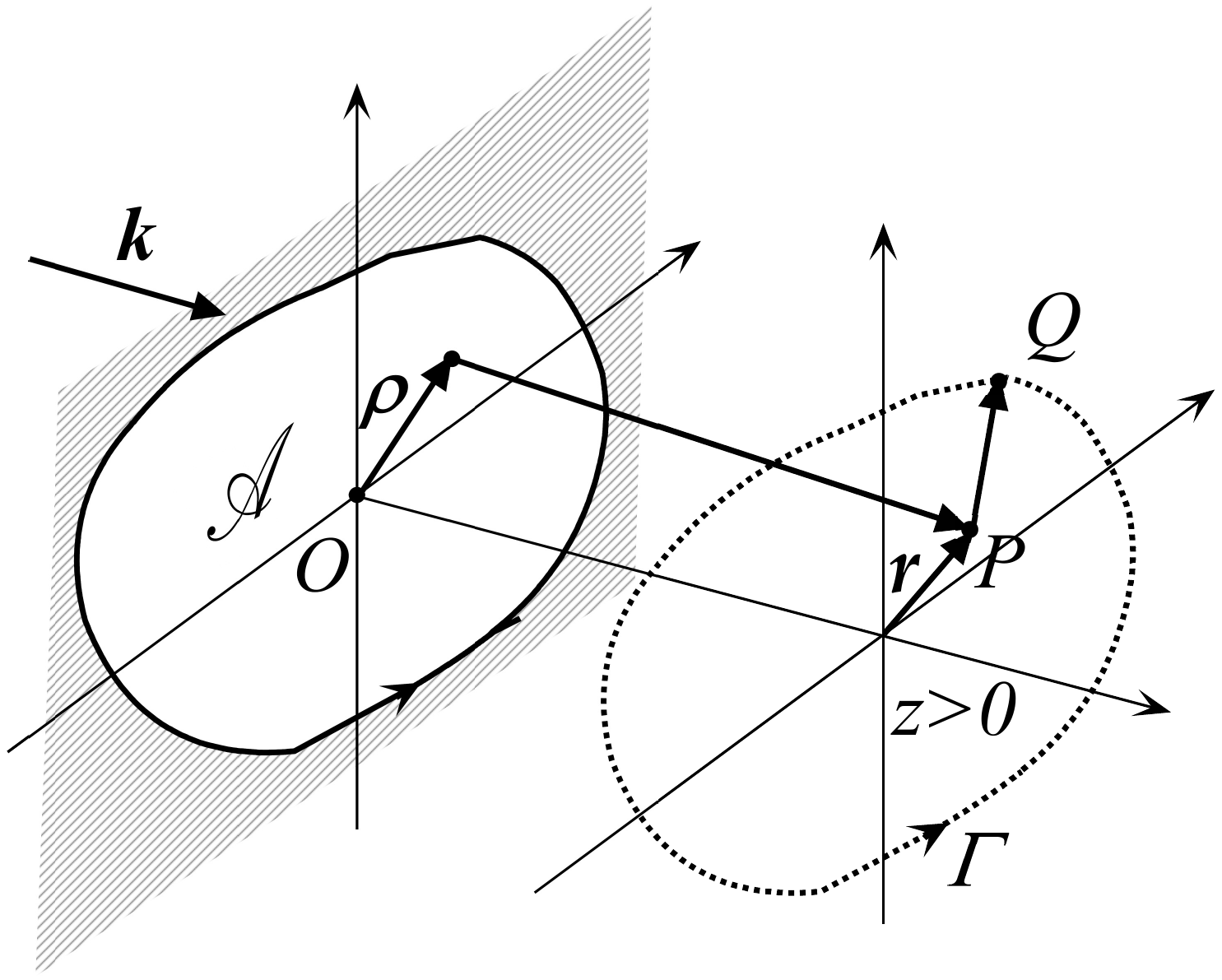}}
\caption{Plane wave sharp-edge diffraction: geometry of  the problem.}
\label{Fig:Fresnel.1}
\end{figure}

Consider a monochromatic plane wave of wavelength $\lambda$ orthogonally  impinging  onto an opaque planar screen 
having a sharp-edge aperture $\mathcal{A}$, as shown in~Fig.~\ref{Fig:Fresnel.1}. 
A unitary (in suitable units) amplitude of the incident wave will be assumed henceforth. 

The paraxially diffracted wavefield at the observation plane $z>0$, say $\psi(\boldsymbol{r};z)$, can be expressed through the following decomposition formula of Fresnel's integral~\cite{Borghi/2015,Borghi/2016}: 
\begin{equation}
\label{Eq:FresnelPropagatorConvolution.3.New.0.1}
\begin{array}{l}
\displaystyle
\psi(\boldsymbol{r};z)\,=\,
-\frac{\mathrm{i}\,k}{2\pi\,z}\,
\int_{\boldsymbol{\mathcal{A}}}\,
\mathrm{d}^2\rho\,
\exp\left(\frac{\mathrm{i}k}{2z}\,|\boldsymbol{r}-\boldsymbol{\rho}|^2\right)\,=\, \\
\,=\,
\psi_G(\boldsymbol{r})\,+\, \psi_{\rm BDW}(\boldsymbol{r};z)\,,
\end{array}
\end{equation}
where an overall phase factor $\exp(\mathrm{i}kz)$ has tacitly been omitted.
Functions $\psi_G(\boldsymbol{r})$ and $\psi_{\rm BDW}(\boldsymbol{r};z)$
are customarily referred to as the \emph{geometrical field} and the \emph{BDW field}, respectively.
Their mathematical definition is 
\begin{equation}
\label{Eq:FresnelPropagatorConvolution.3.New.2}
\begin{array}{l}
\displaystyle
\psi_G\,=\,\frac 1{2\pi}\,\oint_{\Gamma}\,\mathrm{d}\varphi\,,
\end{array}
\end{equation}
and
\begin{equation}
\label{Eq:FresnelPropagatorConvolution.3.New.3}
\begin{array}{l}
\displaystyle
\psi_{\rm BDW}
\,=\,-\frac{1}{2\pi}\,\oint_\Gamma\,\mathrm{d}\varphi\,
\exp\left(\frac{\mathrm{i}U}{2}\,R^2\right)\,,
\end{array}
\end{equation}
{where $\Gamma=\partial \mathcal{A}$ denotes the geometrical projection of the aperture boundary 
across the observation plane, $\boldsymbol{R}=\overrightarrow{PQ}$, with $Q$ being a typical point across $\Gamma$
(see again~Fig.~\ref{Fig:Fresnel.1}).
A polar coordinate reference frame $(R,\varphi)$, centered at the observation point $P$, has been introduced, sa shown in 
Fig.~\ref{Fig:BDW.2}.
Moreover, the dimensionless parameter $U=k\ell^{2}/z$ will be  identified as the Fresnel number,
the parameter $\ell$ being a sort of ``natural'' unit length characteristic of the aperture $\mathcal{A}$.
In this way, also the vectorial quantity $\boldsymbol{R}$ into Eqs.~(\ref{Eq:FresnelPropagatorConvolution.3.New.4}) and~(\ref{Eq:FresnelPropagatorConvolution.3.New.5})
turns out to be dimensionless too. 
\begin{figure}[!ht]
\centerline{\includegraphics[width=4.5cm,angle=-0]{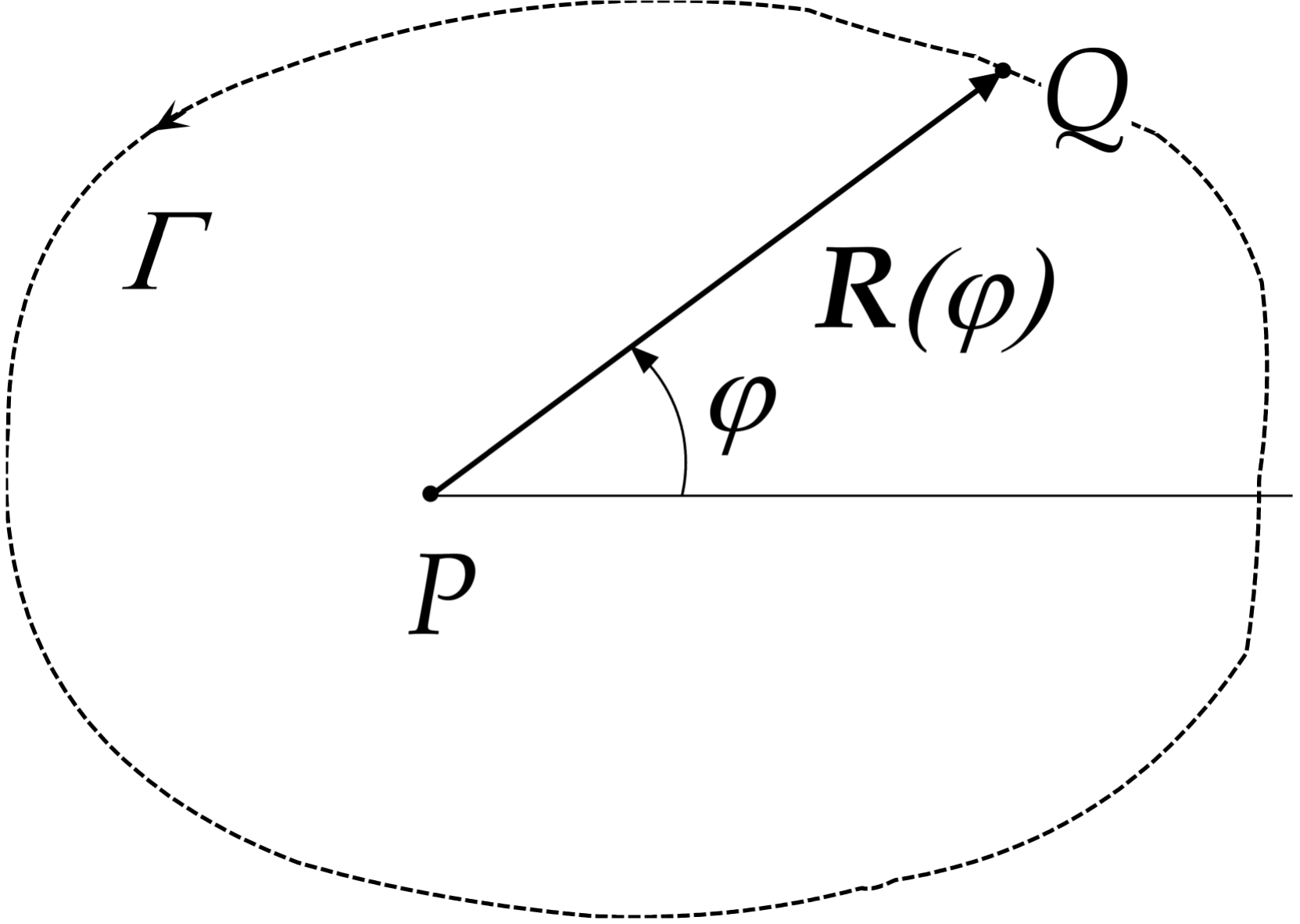}}
\caption{Polar reference frame for  evaluating paraxial integrals in 
Eqs.~(\ref{Eq:FresnelPropagatorConvolution.3.New.2})
and~(\ref{Eq:FresnelPropagatorConvolution.3.New.3}).}
\label{Fig:BDW.2}
\end{figure}

The numerical evaluation of integrals in Eqs.~(\ref{Eq:FresnelPropagatorConvolution.3.New.2}) and~(\ref{Eq:FresnelPropagatorConvolution.3.New.3}) can be done on introducing a suitable parametrization, say $Q=Q(t)$, of the boundary $\Gamma$,  
where $t$ denotes a real parameter ranging within a given interval $\mathcal{I}$. 
Then, Eqs.~(\ref{Eq:FresnelPropagatorConvolution.3.New.2}) and~(\ref{Eq:FresnelPropagatorConvolution.3.New.3})
become
\begin{equation}
\label{Eq:FresnelPropagatorConvolution.3.New.4}
\begin{array}{l}
\displaystyle
\psi_{\rm G}(\boldsymbol{r})\,=\,\frac{1}{2\pi}\oint_\mathcal{I}\,\mathrm{d} t\, 
\dfrac{\boldsymbol{R}\times\dot{\boldsymbol{R}}}
{\boldsymbol{R}\cdot\boldsymbol{R}}\,,
\end{array}
\end{equation}
and
\begin{equation}
\label{Eq:FresnelPropagatorConvolution.3.New.5}
\begin{array}{l}
\displaystyle
\psi_{\rm BDW}(\boldsymbol{r};z)\,=\,-\frac{1}{2\pi}\oint_\mathcal{I}\,\mathrm{d} t\, 
\dfrac{\boldsymbol{R}\times\dot{\boldsymbol{R}}}
{\boldsymbol{R}\cdot\boldsymbol{R}}\,
\exp\left(\frac{\mathrm{i}U}{2}\,\boldsymbol{R}\cdot\boldsymbol{R}\right)\,,
\end{array}
\end{equation}
respectively, where $\dot{\boldsymbol{R}}$ denotes the derivative of $\boldsymbol{R}$ with respect to  $t$ and
the cross product must be intended as the sole $z$-component, being both vectors $\boldsymbol{R}$ and 
$\dot{\boldsymbol{R}}$ purely transverse. 
In particular, $\psi_{\rm G}(\boldsymbol{r})$  coincides with the
\emph{characteristic function} of the aperture $\mathcal{A}$, i.e.,
\begin{equation}
\label{Eq:FresnelPropagatorConvolution.3.New.3.1}
\begin{array}{l}
\displaystyle
\psi_{G}(\boldsymbol{r})\,=\, 
\left\{
\begin{array}{lr}
1 & \boldsymbol{r} \in \mathcal{A}\,,\\
0 & \boldsymbol{r} \notin \mathcal{A}\,.
\end{array}
\right.
\end{array}
\end{equation}
%
The BDW field $\psi_{\rm BDW}$ represents the Young boundary wave.
It is worth stressing that the decomposition formula~(\ref{Eq:FresnelPropagatorConvolution.3.New.0.1}) is an 
\emph{exact} result.

In~\cite{Borghi/2015,Borghi/2016}, it was shown  how the BDW field~(\ref{Eq:FresnelPropagatorConvolution.3.New.3}) can be estimated, for nonsmall  $U$, through a nontrivial asymptotic analysis based on the method of stationary phase~\cite{Born/Wolf/1999}. 
In particular, a pivotal role is played by the saddles of the phase integral in Eqs.~(\ref{Eq:FresnelPropagatorConvolution.3.New.3})
and~(\ref{Eq:FresnelPropagatorConvolution.3.New.5}), which satisfy the following equation:
\begin{equation}
\label{Eq:FresnelPropagatorConvolution.3.New.3.Revised.1}
\begin{array}{l}
\dfrac{\mathrm{d}R^2}{\mathrm{d}\varphi}\,=\,0\,,
\end{array}
\end{equation}
or equivalently, 
\begin{equation}
\label{Eq:SaddleEquation}
\begin{array}{l}
\displaystyle
\boldsymbol{R}(t)\,\cdot\,\dot{\boldsymbol{R}}(t)\,=\,0\,.
\end{array}
\end{equation}
%
From  Eq.~(\ref{Eq:SaddleEquation}), a sound geometrical interpretation of saddles immediately follows: 
they correspond to the orthogonal projections of the observation point $P$ onto the  geometrical shadow 
boundary $\Gamma$, as skecthed in~Fig.~\ref{Fig:BDW.0}.
\begin{figure}[!ht]
\centerline{\includegraphics[width=3.5cm,angle=-90]{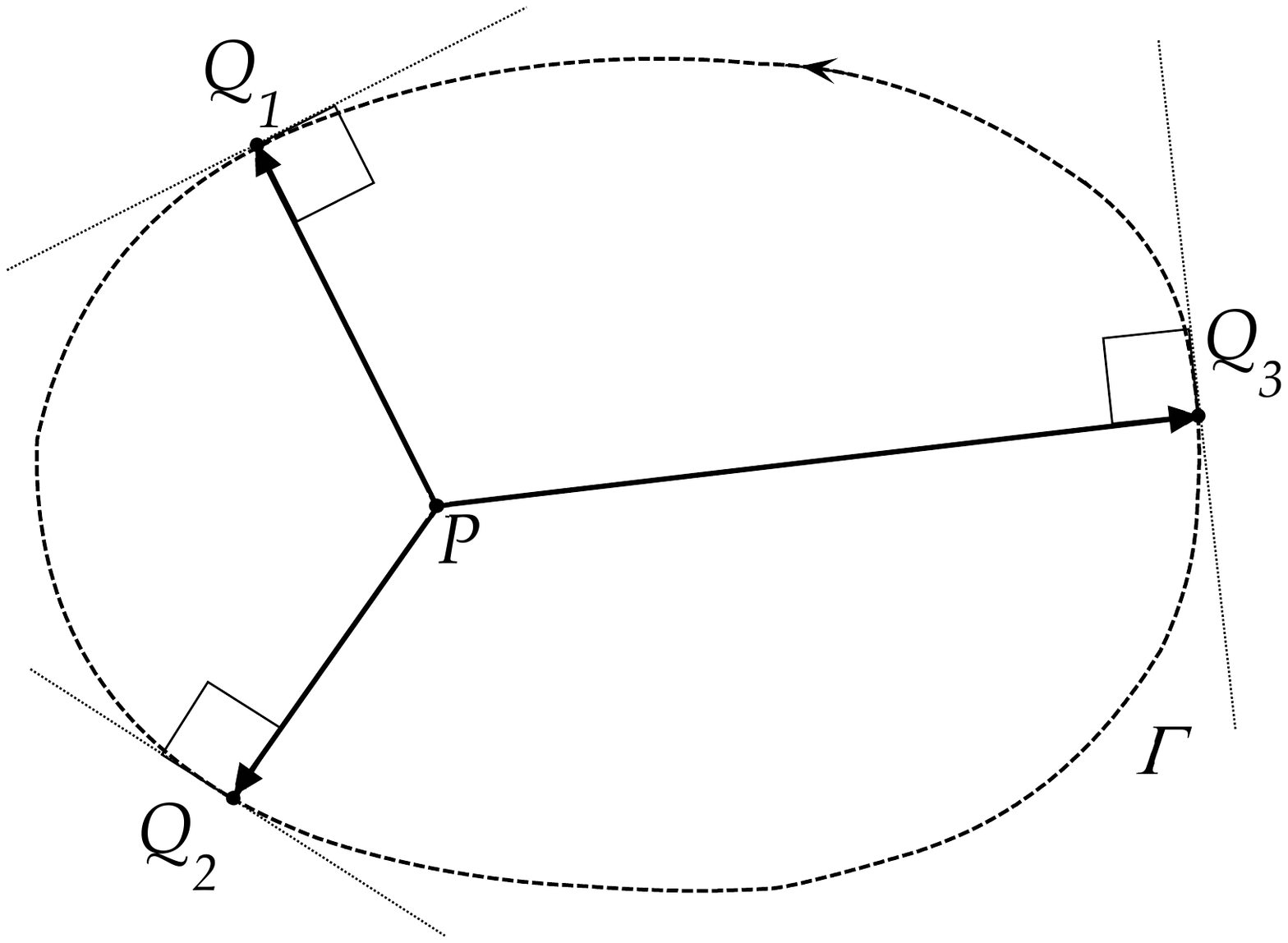}}
\caption{Geometrical interpretation of Eq.~(\ref{Eq:SaddleEquation}).}
\label{Fig:BDW.0}
\end{figure}
The next step is to analyse saddle's dynamics on letting the point $P$ free to move across the observation plane.
Then, not only  saddles move across $\Gamma$, but they can disappear or even suddenly appear on it. 
The birth and the death of saddles correspond to the fact that, on varying the position of $P$,  some of the solutions of Eq.~(\ref{Eq:SaddleEquation}) becomes real or complex, respectively. 
Saddles corresponding to {\em real} solutions of Eq.~(\ref{Eq:SaddleEquation}) are called {\em contributive saddles}.
The fact that an integer quantity  (the number of contributive saddles) be function of a continuous variable 
(the observation point position), unavoidably implies the  presence of discontinuous changes of the former induced by small variations of the latter. Such abrupt changes do occur when two or more contributive saddles coalesce, as well as when some contributive saddle appears on 
$\Gamma$.
In the language of catastrophe optics, saddle coalescing phenomena must be ascribed to the presence of caustics~\cite{Berry/Upstill/1980} which, in the case of plane wave sharp-edge diffraction, are arranged on the geometrical {\em evolute} 
of the aperture boundary $\Gamma$~\cite{Borghi/2015}.

Accordingly, our main problem to be solved is how to ``dress'' the evolute of $\Gamma$ with its proper diffraction patterns, depending on its local topological features. 
This is the mathematical translation of Nye's phrase ``adding wave optics to geometrical optics''.

Sections \ref{Sec:UniformApproximations} and \ref{Sec:Dressing} represent the core of the present paper. There, the dressing problem will be {exactly} solved for 
the cusp ($\mathcal{A}_3$). A single but significant example of application of our algorithm will then be carried out
in Sec. \ref{Sec:NumericalResults}. 

\section{Dressing the cusp: the Catastrophe optics representation of the field}
\label{Sec:UniformApproximations}

\subsection{Preliminaries}
\label{Susec:Preliminari}


Although the dressing problem has already been solved in the past for the fold catastrophe, it is worth reconsidering it in the light of the new algorithm here proposed, in order to help readers to familiarize with it. 
Without loosing generality, we shall refer to a parabolic aperture, similarly as done in ~\cite{Borghi/2016}. 
It represents a very interesting scenario, also from a didactical point of view: first, the number of contributive saddles does not exceed three. Moreover, the evolute of $\Gamma$ consists of two symmetrical fold caustics joined at the parabola focus $F$ to form a single cusp. This is shown in~Fig.~\ref{Fig:ParabolaExample.1}, where the dashed curve represents the evolute of the aperture boundary $\Gamma$ (solid curve).
\begin{figure}[!ht]
\centerline{\includegraphics[width=5cm,angle=-0]{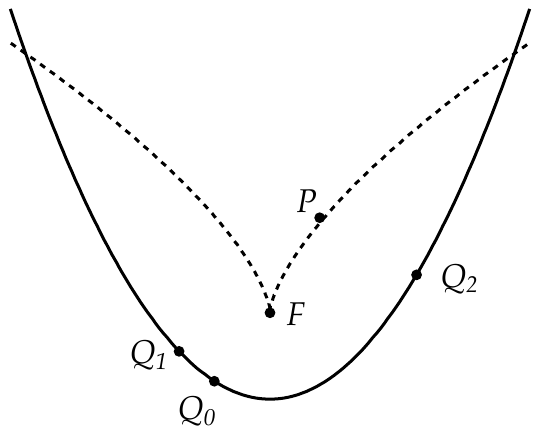}}
\caption{Geometry for studying the plane-wave diffraction from sharp-edge apertures/obstacles. 
Solid curve:  aperture's boundary $\Gamma$. Dashed curve: $\Gamma$'s  evolute. When the observation point $P$ is inside
the cusp, all three saddles $Q_0$,  $Q_1$, and $Q_2$ do contribute to the BDW field.
When $P$ approaches the fold, sufficiently far from $F$, the two saddles $Q_0$ and $Q_1$ are going to coalesce, while $Q_2$ still contributes.
When $P$ approaches $F$, all three saddles are going to coalesce.}
\label{Fig:ParabolaExample.1}
\end{figure}

In other words, parabola represents an ideal scenario for studying the dressing problem of fold and cusp catastrophes.
When the observation point $P$ remains inside the cusp, all three saddles, say $Q_0$,  $Q_1$, and $Q_2$, do 
contribute to the BDW field. But as soon as $P$ approaches the fold (being it sufficiently far from $F$), saddles $Q_0$ and $Q_1$ get closer and closer, up to coalesce ($Q_2$ always continues to contribute).
If, however, $P$ approaches $F$, all three saddles are going to coalesce.
%
%
%
%

\subsection{Catastrophe optics representation of the field near a fold caustic}
\label{Subsec:2Saddeles}

Consider again the situation depicted in~Fig.~\ref{Fig:ParabolaExample.1}, i.e., an observation point is approaching a fold caustic.
We are interested in estimating the asymptotic contribution to the diffracted wavefield $\psi_{\rm BDW}$ 
in Eq.~(\ref{Eq:FresnelPropagatorConvolution.3.New.3}), coming from the {\em pair} $(Q_0, Q_1)$ rather than $Q_0$ and $Q_1$ separately. 
In a seminal paper, Chester~\emph{et al.}~\cite{Chester/Friedman/Ursell/1957} first proposed a general 
procedure to build up such uniform approximations for two coalescing saddles.  
For readers' convenience, this procedures will now be recalled.
The main idea consists in studying the {\em local behaviour} of the function
$R^2(\varphi)$ in the neighborhood of the coalescing point, which resembles that of a \emph{cubic} $\varphi$-polynomial~\cite{Borghi/2016}. Then a new variable, say $\xi$, is first introduced in the diffraction integral~(\ref{Eq:FresnelPropagatorConvolution.3.New.3}), in order for $R^2$ to be approximated, in the neighborhood of the coalescing point, as follows:
\begin{equation}
\label{Eq:Pearcey.1}
\begin{array}{l}
\displaystyle
R^2\,=\,-\frac{\xi^3}3\,+\,\zeta\,{\xi}\,+\,A\,,
\end{array}
\end{equation}
where the {\em real} parameters $(A,\zeta)$ constitute the required ``fold dress,'' as we shall see in a moment.
The subsequent step is to insert from Eq.~(\ref{Eq:Pearcey.1}) into Eq.~(\ref{Eq:FresnelPropagatorConvolution.3.New.3}), which leads to the following estimate of the contribution to $\psi_{\rm BDW}$ coming from the coalescing pair $(Q_0$, $Q_1$):
\begin{equation}
\label{Eq:Pearcey.1.3}
\begin{array}{l}
\displaystyle
-\frac{\exp\left(\mathrm{i}\displaystyle\frac U2\,A\right)}{2\pi}\,
\int_{-\infty}^{+\infty}\,
\mathrm{d}\xi\,\exp\left[-\mathrm{i}\frac U2\left(\frac{\xi^3}3-\zeta\,{\xi}\right)\right]\,
\frac{\mathrm{d}\varphi}{\mathrm{d}\xi}\,,\qquad\qquad U \gg 1\,,
\end{array}
\end{equation}
where the integration limits have been extended to infinite for simplicity (when $U\gg 1$, this is a good approximation).
The integral can be simplified on introducing the power series representation of the factor
${\mathrm{d}\varphi}/{\mathrm{d}\xi}$,
\begin{equation}
\label{Eq:Pearcey.4}
\begin{array}{l}
\displaystyle
\frac{\mathrm{d}\varphi}{\mathrm{d}\xi}\,=\,C_0\,+\,C_1\,\xi\,+\,\ldots\,,
\end{array}
\end{equation}
which, once substituted into Eq.~(\ref{Eq:Pearcey.1.3}), after simple algebra leads to the following 
asymptotics of the 2-saddle contribution~\cite{Borghi/2016}:
\begin{equation}
\label{Eq:Pearcey.1.3.1.4.2}
\begin{array}{l}
\displaystyle
-\left(\dfrac 2U\right)^{1/3}\,\exp\left(\mathrm{i}\dfrac{U}2\,A\right)
\left\{
C_0\,\mathrm{Ai}\left[-\zeta\left(\dfrac U2\right)^{\frac 23}\right]\right.\\
\\
\left. +
\mathrm{i}\,C_1\left(\dfrac 2U\right)^{\frac 13}\,\mathrm{Ai}'\left[-\zeta\left(\dfrac U2\right)^{\frac 23}\right]\,+\,\ldots
\right\}\,.
\end{array}
\end{equation}
In the following, for the sake of simplicity, only the first term of the series will be retained, while the value of $C_0$
will be approximately set to the unity (again, in the limit $U\gg 1$, this turns out to be a good approximation). 
Then, our estimate of the contribution to the BDW field 
coming from a {\em pair} of coalescing point $(Q_0, Q_1)$ turns out to be:
\begin{equation}
\label{Eq:2SaddleContribution}
\begin{array}{l}
\displaystyle
\mathrm{2\,\,saddle\,\,contribution:}\,\\
-\left(\dfrac 2U\right)^{1/3}\,
\exp\left(\mathrm{i}\dfrac{U}2\,A\right)\,\mathrm{Ai}\left[-\zeta\left(\dfrac U2\right)^{\frac 23}\right]\,,
\end{array}
\end{equation}
As said before, the ``fold dress'' is  represented 
by the real pair $(A,\,\zeta)$, 
which 
must (and can be) extracted from the sole geometrical features of the aperture boundary and of its evolute.
%

\subsection{Catastrophe optics representation of the field close to a cusp}
\label{Subsec:3Saddeles}

In presence of three coalescing saddles, things are considerably much more complicated. 
However, the procedure previously illustrated can easily be extended to deal with the present case. 
To this end, it is worth referring to~Fig.~\ref{Fig:ParabolaExample.2}, where now the observation point $P$ is placed sufficiently close to parabola's focus $F$, still inside the cuspoidal region.
\begin{figure}[!ht]
\centerline{\includegraphics[width=5cm,angle=-0]{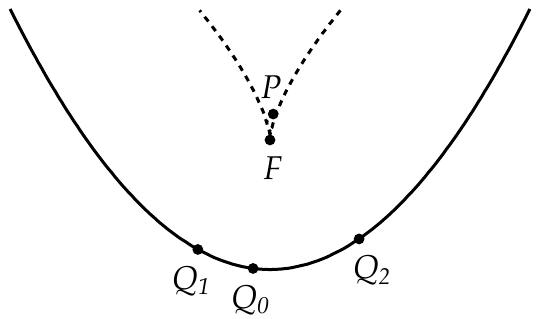}}
\caption{
The same as in~Fig.~\ref{Fig:ParabolaExample.1}, but in the case when all saddles $Q_0$, $Q_1$, and $Q_2$ are going to coalesce.}
\label{Fig:ParabolaExample.2}
\end{figure}

Differently from what happened in~Fig.~\ref{Fig:ParabolaExample.1},  all saddles $Q_0$, $Q_1$, and $Q_2$ are now placed in such a way that it would be impossible to exclude one of them from the approximation. In other words, we need to estimate the contribution to the BDW field coming from the {\em whole} triplet $(Q_0, Q_1,Q_2)$.
First of all, the change of variable $\varphi\to \xi$ has to be chosen in such a way that the function $R^2$ takes on the form, in the neighbourhood of the coalescing point, of a \emph{quartic} $\xi$-polynomial, i.e., 
\begin{equation}
\label{Eq:ThreeSaddles}
\begin{array}{l}
\displaystyle
R^2\,=\,\frac{\xi^4}4\,-\,\zeta\,\frac{\xi^2}2\,-\,\eta\,\xi\,+\,A\,,
\end{array}
\end{equation}
where the new parameter $\eta$ has been introduced.
On proceeding similarly as done in the previous section, the contribution to the BDW integral coming from  $(Q_0, Q_1,Q_2)$
will be written as
\begin{equation}
\label{Eq:Pearcey.1.3Bis}
\begin{array}{l}
\displaystyle
-\frac{\exp\left(\mathrm{i}\displaystyle\frac U2\,A\right)}{2\pi}\,
\int_{-\infty}^{+\infty}\,
\mathrm{d}\xi\,\exp\left[\mathrm{i}\frac U2\left(\frac{\xi^4}4\,-\,\zeta\,\frac{\xi^2}2\,-\,\eta\,\xi\right)\right]\,\frac{\mathrm{d}\varphi}{\mathrm{d}\xi}\,.
\end{array}
\end{equation}
Then, on again substituting from Eq.~(\ref{Eq:Pearcey.4}), truncated at the first term, into Eq.~(\ref{Eq:Pearcey.1.3Bis}) and on further doing the change of variable $\xi\,=\, t (U/8)^{-1/4}$, long but straightforward leads to 
\begin{equation}
\label{Eq:3SaddleContribution}
\begin{array}{l}
\displaystyle
\mathrm{3-saddle\,contribution:}\,\\
-\dfrac {1}{\pi\,(2U)^{1/4}}\,\exp\left(\mathrm{i}\dfrac{U}2A\right)\,\mathrm{Pe}\left[-\eta\left(\dfrac {U^3}2\right)^{1/4},-\zeta\left(\dfrac {U}2\right)^{1/2}\right]\,,
\end{array}
\end{equation}
where the function $\mathrm{Pe}(s_1,s_2)$, defined as
\begin{equation}
\label{Eq:Pearcey.0}
\begin{array}{l}
\displaystyle
\mathrm{Pe}(s_1,s_2)\,=\,
\int_{-\infty}^{+\infty}\,
\mathrm{d} t\,\exp\left[\mathrm{i}\left({t^4}\,+\,s_2\,{t^2}\,+\,s_1\,{t}\right)\right]\,,
\end{array}
\end{equation}
is the so-called Pearcey function~\cite{Pearcey/1946}.
The ``cusp dress'' is now represented by the real triplet 
$(A,\,\zeta,\,\eta)$ which must again be extracted from geometry.
This will be the subject of the next section, which contains  the very new contribution of the present paper.

\section{Dressing the cusp: the algorithm}
\label{Sec:Dressing}

\subsection{Preliminaries}
\label{Subsec:Parameters:Preliminaries}

After the analytical structure of the diffracted field in the neighborhood of the geometrical singularities (fold and cusp) has been established, it remains to solve the dressing problem: to find the arguments of the functions comparing into Eqs.~(\ref{Eq:2SaddleContribution}) and~(\ref{Eq:3SaddleContribution}),
i.e., 
$A$, $\zeta$, and $\eta$ (for the sole cusp). As anticipated, such values depend only on the geometry of the diffraction problem, namely, on the boundary shape, the evolute shape, and on the observation point location.
Remarkably, the Fresnel number, which contains the ondulatory information (through the 
wavelength), appears to be nothing but a scale factor (via its powers) of the diffraction catastrophe arguments. 
In Sec.~\ref{Sec:NumericalResults}, such an important scaling law will be numerically validated. Now, the dressing problem
will be tackled and solved for both fold and cusp.
As far as the former is concerned, it must be said that it has already been solved in~\cite{Borghi/2016}
on using an elementary approach. But for three coalescing saddles, the mathematical difficulties appeared so severe to discourage any attempt of analytical solution. The main technical reason is the presence of a cubic algebraic equation which, if were explicitly solved, would dramatically increase the complexity of the subsequent steps.

The main result of the present paper is an algorithm for the dressing problem to be solved through simple algebraic formulas, without using explicit solutions of algebraic equations. Surprisingly enough, such a task can be accomplished by using a beautiful mathematical theorem attributed to Isaac Newton during his {\em annus mirabilis} 1666. 
Although explicit solutions for $A$ 
and $\zeta$ have already been found in~\cite{Borghi/2016} for the case of two coalescing saddles, it is worth retrieving 
the same results in the light of Newton's identities, before tackling the more 
challenging problem of three coalescing saddles. 

\subsection{Dressing folds}
\label{Subsec:Parameters:2Saddles}

The idea is 
to express the pair $(A,\zeta)$ in terms of the {values} 
attained by the function $R^2$ at the \emph{sole coalescing saddles}. In the following, such quantities will be denoted $R^2_0=\overline{PQ}_0^2$ and $R^2_1=\overline{PQ}_1^2$.
Saddles $(Q_0,Q_1)$ correspond to the solutions of $\mathrm{d}R^2/{\rm d}\xi=0$ which, on taking Eq.~(\ref{Eq:Pearcey.1}) into account, coincide with the roots, say $(\xi_0,\xi_1)$, of the following 
{quadratic} equation: 
\begin{equation}
\label{Eq:Parameters.1}
\begin{array}{l}
\displaystyle
{\xi^2}\,-\,\zeta\,=\,0\,.
\end{array}
\end{equation}
Once roots are substituted again into Eq.~(\ref{Eq:Pearcey.1}), the following system is then obtained:
\begin{equation}
\label{Eq:Parameters.2}
\begin{array}{l}
\displaystyle
R_k^2\,=\,-\frac{\xi_k^3}3\,+\,\zeta\,{\xi_k}\,+\,A\,,\qquad\qquad k\,=\,0,1,
\end{array}
\end{equation}
which must be solved in terms of $\zeta$ and $A$. Before continuing, 
it is worth stressing how Eq.~(\ref{Eq:Parameters.2}) be mathematically much more complex as it could be appear at first sight. While the left side contains quantities depending only on the relative position of the observation point $P$ with respect to the aperture boundary,  the right side is linear with respect to $A$, but genuinely nonlinear with respect to the other unknown $\zeta$, once the saddle equation~(\ref{Eq:Parameters.1}) would be explicitly solved.
However, in this particular case (two coalescing saddles), system~(\ref{Eq:Parameters.2}) can be solved in an elementary way~\cite{Borghi/2016}, due to the fact that the roots of Eq.~(\ref{Eq:Parameters.1})  are simply $\pm\,\sqrt\zeta$.

The new approach proposed here, which is based on Newton's identities, will now be tested on this particular case, also to help readers to familiarize with it.  
To this end, quantities $\sigma_n\,=\,\displaystyle\sum_k\,R^n_k$  ($n=2,3,\ldots$),  are first introduced. 
Here the symbol $\displaystyle\sum_k\cdot$ denotes summation with respect the values of $k$ 
corresponding to all roots. In the present case, it should be meant $\displaystyle\sum_k\cdot=\displaystyle\sum_{k=0}^1\cdot$

Then, on summing side by side Eq.~(\ref{Eq:Parameters.2}), we have
\begin{equation}
\label{Eq:Parameters.3}
\begin{array}{l}
\displaystyle
\sigma_2\,=\,
-\dfrac{1}3\,\sum_k\xi_k^3\,+\,\zeta\,\sum_k{\xi_k}\,+\,2\,A\,.
\end{array}
\end{equation}
Newton's identities, once applied to the roots $\xi_k$ of Eq.~(\ref{Eq:Parameters.1}), yield
\begin{equation}
\label{Eq:Newton.1}
\left\{
\begin{array}{l}
\displaystyle
\sum_k\xi_k=0\,,\\
\displaystyle
\sum_k\xi_k^3=0\,,
\end{array}
\right.
\end{equation}
so that Eq.~(\ref{Eq:Parameters.3}) leads to $\sigma_2=2A$, i.e.,
\begin{equation}
\label{Eq:Parameters.3.1}
\begin{array}{l}
\displaystyle
A\,=\,\frac{\sigma_2}2\,=\,\dfrac{R^2_0+R^2_1}2\,.
\end{array}
\end{equation}
To retrieve the parameter $\zeta$, both sides of Eq.~(\ref{Eq:Parameters.2}) are first squared, i.e.,
\begin{equation}
\label{Eq:Parameters.4}
\begin{array}{l}
\displaystyle
R_k^4\,=\,
\frac{\xi_k^6}9\,+\,\zeta^2\,{\xi^2_k}\,+\,A^2\,
\,-\,\frac 23\,\zeta\,\xi^4_k\,-\,\frac 23\,A\,\xi^2_k\,+\,2\zeta A\xi_k\,,\qquad k=0,1\,.
\end{array}
\end{equation}
In order to proceed as before, the list of Newton's identities must be updated as follows:
\begin{equation}
\label{Eq:Newton.2}
\left\{
\begin{array}{l}
\displaystyle
\sum_k\xi^2_k=2\zeta\,,\\
\displaystyle
\sum_k\xi^4_k=2\zeta^2\,,\\
\displaystyle
\sum_k\xi^6_k=2\zeta^3\,.
\end{array}
\right.
\end{equation}
On summing again both sides of Eq.~(\ref{Eq:Parameters.4}) and on taking Eqs.~(\ref{Eq:Newton.1}) and~(\ref{Eq:Newton.2}) into account, long but in principle straightforward algebra gives
\begin{equation}
\label{Eq:Parameters.5}
\begin{array}{l}
\displaystyle
\sigma_4\,=\,\sum_kR_k^4\,=\,2\,A^2\,+\,\frac{8}9\,\zeta^3\,.
\end{array}
\end{equation}
Last step consists in substituting from Eq.~(\ref{Eq:Parameters.3.1}) into Eq.~(\ref{Eq:Parameters.5}), 
so that
\begin{equation}
\label{Eq:Parameters.6}
\begin{array}{l}
\displaystyle
\zeta^3\,=\,\frac 9{16}\,(2\sigma_4\,-\,\sigma^2_2)\,, 
\end{array}
\end{equation}
which complete the solution. 
In particular, by noting that
\begin{equation}
\label{Eq:Parameters.6.1}
\begin{array}{l}
\displaystyle
2\sigma_4\,-\,\sigma^2_2\,=\,
2(R^4_0+R^4_1)-(R^2_0+R^2_1)^2\,=\,(R^2_0-R^2_1)^2\,,
\end{array}
\end{equation}
the real solution of Eq.~(\ref{Eq:Parameters.6}) can be explicited as follows:
\begin{equation}
\label{Eq:Parameters.6.2}
\begin{array}{l}
\displaystyle
\zeta\,=\,\left[\dfrac 34(R^2_0-R^2_1)\right]^{2/3}\,.
\end{array}
\end{equation}
Equations~(\ref{Eq:Parameters.3.1}) and~(\ref{Eq:Parameters.6.2}) coincide with Eq.~(10) of~\cite{Borghi/2016}.
The new approach based on Newton's identities seems to work well for two coalescing saddles. 
When both $Q_0$ and $Q_1$ do contribute to the diffracted field (i.e., when the observation point $P$ has not yet reached the fold caustic), $R^2_0$ and $R^2_1$ are real and positive quantities, so that $\zeta>0$.
Coalescence occurs exactly at  $\zeta=0$, while on further letting $P$ to leave caustics, quantities $R^2_k$ become complex 
conjugates and Eq.~(\ref{Eq:Parameters.6}) will predict $\zeta$ to attains {negative real} values.

\subsection{Dressing cusps}
\label{Subsec:Parameters:3Saddles}

In the case of three coalescing saddles, equation $\mathrm{d}R^2/{\rm d}\xi=0$
becomes, after taking Eq.~(\ref{Eq:ThreeSaddles}) into account, 
\begin{equation}
\label{Eq:Parameters.13}
\begin{array}{l}
\displaystyle
{\xi^3}\,-\,\zeta\,\xi\,-\,\eta\,=\,0\,.
\end{array}
\end{equation}
It is important to put into evidence how, due to the hypothesis $(\zeta,\eta)\in\mathbb{R}^2$, only two scenarios are possible.
In the first one, all roots of Eq.~(\ref{Eq:Parameters.13}) are real, which corresponds to have three contributive saddles.  
In the second scenario, only one root is real, the other two being complex conjugate. In such scenario two saddles coalesced and then disappeared, leaving only a single contributive saddle. 
Accordingly, the system analogue to that in Eq.~(\ref{Eq:Parameters.2}) is the following:
\begin{equation}
\label{Eq:Parameters.14}
\begin{array}{l}
\displaystyle
R_k^2\,=\,\frac{\xi_k^4}4\,-\,\zeta\,\frac{\xi_k^2}2\,-\,\eta\xi_k\,+\,A\,,\qquad\qquad k\,=\,0,1,2\,.
\end{array}
\end{equation}
Differently from what happened in the previous section, 
Eq.~(\ref{Eq:Parameters.14}) cannot be solved for the triple $(A,\zeta,\eta)$ in an elementary way. 
The reason is the nontrivial dependence (through Tartaglia/Cardano's formula) of the roots $(\xi_0,\,\xi_1,\,\xi_2)$ of 
Eq.~(\ref{Eq:Parameters.13}) on the pair $(\zeta,\eta)$. 
Nevertheless, Newton's identities allow the analytical solution of the cusp dressing problem to be found. A mathematical miracle, 
in spite of the premises.
First of all, from Eq.~(\ref{Eq:Parameters.14}) we have
\begin{equation}
\label{Eq:Parameters.14.1}
\begin{array}{l}
\displaystyle
\sigma_2\,=\,\frac{1}4\,\displaystyle\sum_k\xi_k^4\,-\,\frac{\zeta}2\,\displaystyle\sum_k\xi_k^2\,-\,\eta\displaystyle\sum_k\xi_k\,+\,3A\,,\qquad\qquad k\,=\,0,1,2\,,
\end{array}
\end{equation}
where now and henceforth, it should be meant $\displaystyle\sum_k\,=\sum_{k=0}^2$. 

In order to proceed as before, the list of the useful Newton identities for the cubic equation~(\ref{Eq:Parameters.13}) has to
be arranged. To improve paper's readability, such list has been confined into Appendix \ref{App:NonlinearSystem}.
In particular, substitution from Eq.~(\ref{Eq:Parameters.NewtonTheor}) into Eq.~(\ref{Eq:Parameters.14.1}) gives at once
\begin{equation}
\label{Eq:Parameters.14.2}
\begin{array}{l}
\displaystyle
\sigma_2\,=\,3A\,-\,\dfrac{\zeta^2}2\,,
\end{array}
\end{equation}
i.e., the first equation.
Other two equations can be obtained from Eq.~(\ref{Eq:Parameters.14}) simply by looking at the quantities 
$\sigma_4$ and $\sigma_6$. The subsequent algebra, mostly straightforward although somewhat boring, is again 
confined into Appendix~\ref{App:NonlinearSystem}, where it is proved that
\begin{equation}
\label{Eq:Parameters.15}
\left\{
\begin{array}{l}
\displaystyle
\sigma_2\,=\,3A\,-\,\dfrac{\zeta^2}2\,,\\
\\
\sigma_4\,=\,3A^2\,-\,A{\zeta^2}\,+\,\dfrac94\,\zeta\eta^2\,+\,\dfrac{\zeta^4}8\,,\\
\\
\sigma_6\,=\, 3 A^3\,-\,\dfrac 32 A^2\zeta^2\,+\,\dfrac 38 A\zeta^4\,-\,\dfrac{\zeta^6}{32}\,+\,\dfrac{27}{4} A\zeta\eta^2\,\\
\qquad-\,\dfrac{51}{32}\zeta^3\eta^2\,-\,\dfrac{81}{64} \eta^4\,.
\end{array}
\right.
\end{equation}
Equation~(\ref{Eq:Parameters.15}) represents one of the main results of the present paper. 

Before continuing, it is worth checking that, when $P$ coincides with the coalescing point (for instance, the focus $F$ when the aperture coincides with the parabola), the system~(\ref{Eq:Parameters.15}) admits a trivial solution. In fact, we should have 
$R^2_k=R^2_0$ for $k=0,1,2$, so that $\sigma_2=3R^2_0$, $\sigma_4=\sigma_2/3$, and $\sigma_6=\sigma_2/9$.
Accordingly, it is trivial to verify that Eq.~(\ref{Eq:Parameters.15}) is satisfied by $\zeta=\eta=0$ and $A=\sigma_2/3=R^2_0$.
This result is in perfect agreement with the results obtained for the fold caustic, where the couple 
$\left(\zeta=0\,,A={\sigma_2}/2\right)$ occured whenever $P$ coincided with one the infinite coalescing points constituting the fold itself. In the case of cusps, there will be only a {\em finite} number of coalescing points (i.e., the points where two folds join together). So, for example, parabola has only one coalescing point, while ellipses have four.

When $P$ is not a coalescing point, to find a solution of the highly nonlinear system~(\ref{Eq:Parameters.15}) turns out to be a considerably harder job.
However, it is possible to reduce the system to a single algebraic equation for $\zeta$, from which the other two parameters can be
obtained by simple substitution. The, nontrivial, sequence of mathematical steps is detailed in Appendix~\ref{App:SolvingSystem}, where it is proved that Eq.~(\ref{Eq:Parameters.15}) transforms into
\begin{equation}
\label{Eq:Parameters.16}
\left\{
\begin{array}{l}
\displaystyle
27\,\zeta^8\,-\,18\tau_4\zeta^4-4\tau_6\zeta^2-\tau^2_4=0\,,\\
\eta^2\,=\,\dfrac{\tau_4\,-\,\zeta^4}{54\,\zeta}\,,\\
A\,=\,\dfrac{\sigma_2}3\,+\,\dfrac{\zeta^2}6\,,
\end{array}
\right.
\end{equation}
where the quantities $\tau_4$ and $\tau_6$ are defined by
\begin{equation}
\label{Eq:Parameters.17}
\left\{
\begin{array}{l}
\displaystyle
\tau_4\,=\,24\sigma_4-8\sigma_2^2\,,\\
\tau_6\,=\,576\,(\sigma_6\,-\,\sigma_4\sigma_2)\,+\,128\,\sigma^3_2\,.
\end{array}
\right.
\end{equation}
It is worth giving the following explicit expressions of the quantities $\tau_4$ and $\tau_6$ 
in terms of the values of the function $R^2$ at the three saddles:
\begin{equation}
\label{Eq:Parameters.21}
\left\{
\begin{array}{l}
\displaystyle
\tau_4\,=\,8\sum_{i\ne j}\,(R^2_i\,-\,R^2_j)^2\,,
\\
\tau_6\,=\,64\,\\
\times(2R^2_0\,-\,R^2_1\,-\,R^2_2)\,(2R^2_1\,-\,R^2_2\,-\,R^2_0)\,(2R^2_2\,-\,R^2_0\,-\,R^2_1)\,,
\end{array}
\right.
\end{equation}
from which it follows $\tau_4$ to be intrinsically nonnegative.
Nothing can be said about $\tau_6$, except that it is real. 
The eight-order algebraic equation with respect to the variable $\zeta$ represents the key to solve the problem.
In fact, on letting $\chi=\zeta^2$, such equation becomes
\begin{equation}
\label{App:SolvingSystem.5.1}
\begin{array}{l}
\displaystyle
27\,\chi^4\,-\,18\tau_4\chi^2-4\tau_6\chi-\tau^2_4=0\,.
\end{array}
\end{equation}
It is not difficult to prove that Eq.~(\ref{App:SolvingSystem.5.1}) admits only one nonnegative root\cite{Prodanov/2021}, whose analytical expression can be obtained  via Cardano's formula, precisely
\begin{equation}
\label{Eq:Parameters.23}
\zeta^2\,=\,\dfrac{\sqrt{3\tau_4\,-\,\Delta^2\,+\,\dfrac{|\tau_6|}\Delta}\,+\,\Delta\,\mathrm{sgn}(\tau_6)}3\,,
\end{equation}
where the symbol sgn$(\cdot)$ denotes the signum function and the 
the quantity $\Delta$ is defined by
\begin{equation}
\label{Eq:Parameters.23.1}
\Delta=\sqrt{\tau_4\,-\,\sqrt[3]{\tau^3_4\,-\,\dfrac{\tau^2_6}4}}\,.
\end{equation}
Moreover, from Eq.~(\ref{Eq:Parameters.21}) it is possible to prove, with long but straigthforward algebra, the following notable equation:
\begin{equation}
\label{Eq:Parameters.22}
4\tau^3_4\,-\,\tau^2_6\,=\,48^3\,\prod_{i\ne j}\,(R^2_i\,-\,R^2_j)^2\,\ge\,0\,,
\end{equation}
from which it follows that $\Delta \ge 0$.

\section{Numerical Results}
\label{Sec:NumericalResults}

\subsection{Preliminaries}
\label{Subsec:NumericalResults:Preliminaries}

In the present section, a single numerical example will be carried out, in order to show the practical applicability of
the algorithm developed in the previous section.
As anticipated, the sharp-edge diffraction from a parabolic aperture will now be considered, following the path traced in~\cite{Borghi/2016}.
It must be recalled that sharp-edge diffraction from parabolic apertures (obstacles) was experimentally studied since the early days of the so-called geometrical theory of diffraction~\cite{Raman/1919}. In 1922, Coulson and Becknell published a, nearly forgotten, series of beautiful papers~\cite{Coulson/Becknell/1922,Becknell/Coulson/1922b}, in the second of which parabolic opaque planar plates were experimentally studied. From our CO perspective, parabola represents an ideal benchmark for the scopes of the present paper, 
because the number of contributive saddles never exceeds three, regardless the position of $P$. 
In~\cite{Borghi/2016}, the 
dressing problem has been tackled and solved for the parabola fold caustic, according to the results recalled in Sec.~\ref{Subsec:Parameters:2Saddles}. Now, the analytical solution provided in Sec.~\ref{Subsec:Parameters:3Saddles} will be employed to solve the dressing problem for the parabola cusp. As we shall see in a moment,  {\em a priori} unpredictable (at least for me) simple and elegant analytical expressions for $(A,\zeta,\eta)$ will be found. 

Consider again the parabola of~Fig.~\ref{Fig:ParabolaExample.1} and suppose the distance between the focus $F$ and parabola's vertex to be unitary (in suitable units). A Cartesian reference frame $OXY$ is then introduced, with $O\equiv F$ and with the $Y$ axis aligned along the parabola axis. The Cartesian equation of parabola's evolute turns out to be $27X^2=8Y^3$. 
In other words, on denoting $P \equiv (X,Y)$ the Cartesian representation of the observation point, 
the inequality  $27X^2<8Y^3$ implies $P$ to be located {\em inside} the cuspoidal region, whereas for $27X^2>8Y^3$, $P$ is outside of it. The next step is to introduce a parametric representation of parabola, for instance $Q(t)=\left(t,\dfrac{t^2}2-1\right)$ with $t\in \mathbb{R}$, in order to numerically evaluate the BDW field directly via Eq.~(\ref{Eq:FresnelPropagatorConvolution.3.New.5}).
In~Fig.~\ref{Fig:ParabolaIntensityMap}, a 2D map of $|\psi_{\rm BDW}|^2$, evaluated for $U=10^3$, is
plotted within the plane $(X,Y)$. The white dashed  curve represents the parabola evolute. It is worth comparing such figure with the photograph reported in Plate I, No. 2 of~\cite{Becknell/Coulson/1922b}.
\begin{figure}[!ht]
\centerline{\includegraphics[width=4cm,angle=-0]{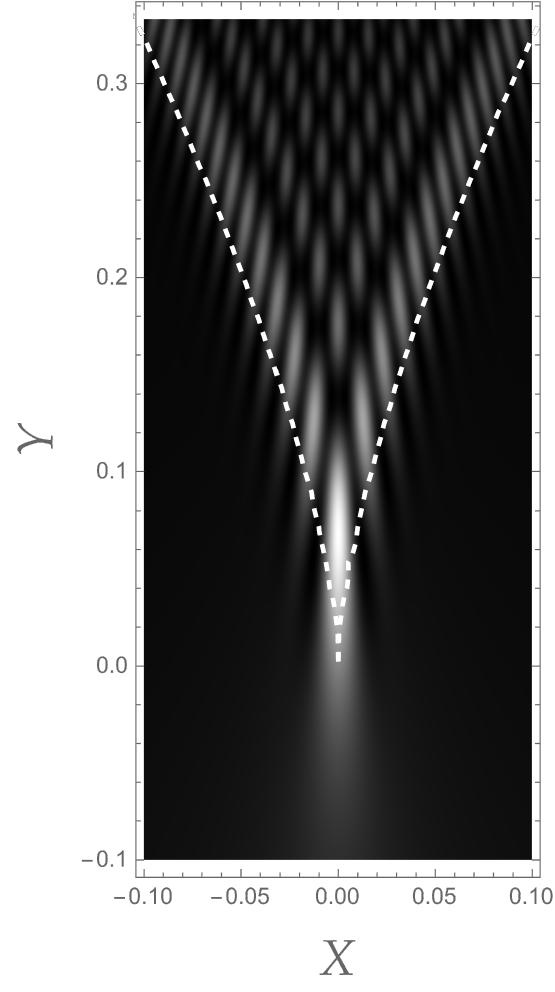}}
\caption{2D map of the intensity distribution of the BDW field
obtained via Eq.~(\ref{Eq:FresnelPropagatorConvolution.3.New.5}), for $U=10^3$.
Parabola's parametrization $Q(t)=\left(t,\dfrac{t^2}2-1\right)$, with $t\in \mathbb{R}$, has been used.
Dashed curve represents parabola's evolute.}
\label{Fig:ParabolaIntensityMap}
\end{figure}
The numerical evaluation of the 2D map, carried out via the Mathematica native command {\tt NIntegrate}, 
did not present any problem, despite the high value of Fresnel's number. As we shall see in a
moment, this will be of great help for the subsequent asymptotic characterization of the BDW field in proximity 
of the focus $F$. Our analysis starts with observation points located
along the parabola axis $Y$. Later, the more general scenario will be studied. 
%
%
In any case, the preliminary task consists in implementing saddle's equation~(\ref{Eq:SaddleEquation}) in terms of the parabola parametrization $Q(t)$. To this end, the Cartesian representation of the position vector $\boldsymbol{R}$ turns out to be $\boldsymbol{R}\,=\,Q(t)-P\,=\,\left(t-X,\dfrac{t^2}2-1-Y\right)$, from which it follows $\dot{\boldsymbol{R}}=\left(1,t\right)$. Finally, Eq.~(\ref{Eq:SaddleEquation}) becomes
\begin{equation}
\label{Eq:SaddleEquationParabola}
\begin{array}{l}
\displaystyle
t^3\,-\,2Y\,t\,-\,2X\,=\,0\,.
\end{array}
\end{equation}

\subsection{Dressing parabola's cusp: on-axis analysis}
\label{Subsec:NumericalResults:Parabola:OnAxis}

As said above, our asymptotic analysis of the BDW field starts on setting $X=0$. Accordingly, the saddles positions can be found by solving 
the algebraic equation $t^3 - 2 Y t =0$, which gives at once $t=0$ and $t=\pm\sqrt{2Y}$. 
Accordingly, parabola's parametrization we obtainleads to
\begin{equation}
\label{Eq:NumericalResults:Parabola.0.1}
\left\{
\begin{array}{l}
\displaystyle
Q_0\equiv \left(0,-1\right)\,,\\
Q_1\equiv \left(-\sqrt{2Y},Y-1\right)\,,\\
Q_2\equiv \left(\sqrt{2Y},Y-1\right)\,.
\end{array}
\right.
\end{equation}
As expected, when $Y>0$ all three saddles do contribute.  At $Y=0$ (i.e., when $P \equiv F$), they 
coalesce into the parabola vertex $ V \equiv Q_0$, which remains the only contributive saddle for $Y<0$.
Due to the symmetry, it also must be expected that $\eta=0$. 
Then, on taking Eq.~(\ref{Eq:NumericalResults:Parabola.0.1}) into account, it is not difficult to prove that
\begin{equation}
\label{Eq:NumericalResults:Parabola.1}
\left\{
\begin{array}{l}
\displaystyle
R^2_0=(Y+1)^2\,,\\
R^2_1\,=\,R^2_2\,=\, 2Y+1\,,
\end{array}
\right.
\end{equation}
which, once substituted into Eq.~(\ref{Eq:Parameters.21}), gives at once $\tau_4=16 Y^4$ and $\tau_6=128 Y^6$, so that $4\tau^3_4-\tau^2_6\equiv 0$. Accordingly, from Eq.~(\ref{Eq:Parameters.23}) we have 
$\zeta^2=4Y^2$ and
finally, after taking Eq.~(\ref{Eq:Parameters.16}) into account,
\begin{equation}
\label{Eq:NumericalResults:Parabola.5.1}
\left\{
\begin{array}{l}
\displaystyle
\zeta(Y)\,=\, 2Y\,,\\
\eta(Y)\,=\,0\,,\\
A(Y)\,=\,(1+Y)^2\,,
\end{array}
\right.
\end{equation}
which gives the complete solution to the on-axis cusp dressing problem.
The mathematical simplicity of Eq.~(\ref{Eq:NumericalResults:Parabola.5.1}) should not surprise, due to the symmetry of the problem. 
It is now worth checking the prediction capabilities of the asymptotics in Eq.~(\ref{Eq:3SaddleContribution}).
To this end, Eq.~(\ref{Eq:NumericalResults:Parabola.5.1}) will be employed
similarly as we did in~\cite{Borghi/2017} as far as the simpler scenario of  two coalescing saddles was concerned.
\begin{figure}[!ht]
\centerline{\includegraphics[width=8cm,angle=-0]{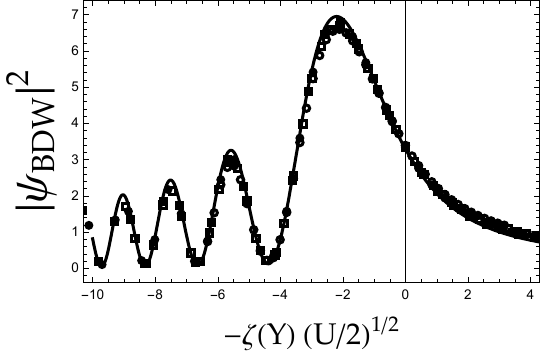}}
\caption{Behavior of the intensity distribution of the BDW field
across the $Y$-axis, obtained via  Eq.~(\ref{Eq:FresnelPropagatorConvolution.3.New.5}), for $U=500$ (open circles), $U=1000$
(open squares)  $U=2000$ (black circles) and $U=3000$ (black squares).
The intensity values  are plotted against the normalized variable $s_2= -\zeta(Y)\,\sqrt{\dfrac U2}$,
where $\zeta(Y)$ is given by Eq.~(\ref{Eq:NumericalResults:Parabola.5.1}). 
The solid curve represents the function $|\mathrm{Pe}(0,s_2)|^2$.}
\label{Fig:PearceyScalingLaw}
\end{figure}
This is shown in~Fig.~\ref{Fig:PearceyScalingLaw}, where values of the BDW intensity $|\psi_{\rm BDW}|^2$, 
numerically computed via Eq.~(\ref{Eq:FresnelPropagatorConvolution.3.New.5}) along the $Y$ axis and suitably normalized by a factor $\pi^2\sqrt{2U}$, are plotted against the scaled variable $s_2=-\zeta(Y)\,\sqrt{\dfrac U2}$, for several values of the Fresnel number, namely $U=500,\,1000,\,2000,\,3000$, with $\zeta(Y)$ being given into Eq.~(\ref{Eq:NumericalResults:Parabola.5.1}).
In the same graph, the function $|\mathrm{Pe}(0, s_2)|^2$ is also plotted as a solid curve, in order to confirm  
the goodness of Eq.~(\ref{Eq:3SaddleContribution}).

\subsection{Dressing parabola's cusp: off-axis case}
\label{Subsec:NumericalResults:Parabola:OffAxis}

Consider now the considerably more complex off-axis case, i.e., $X \ne 0$. 
In order to determine the saddle locations along $\Gamma$, the complete cubic equation~(\ref{Eq:SaddleEquationParabola}) should now be solved. 
Of course, this could be done on using Cardano's formula.
However, in the present case (parabola) it is possible to guess the pair $(\zeta,\eta)$ without
doing any calculation. To this end, it is sufficient to compare Eq.~(\ref{Eq:Parameters.13}) with 
Eq.~(\ref{Eq:SaddleEquationParabola}),  and to note that such equations become formally identical 
simply on letting $\zeta=2Y$ and $\eta=2X$. The third parameter $A$ can be found
by evaluating the quantity $\sigma_2$, what can be done again without solving explicitly 
Eq.~(\ref{Eq:SaddleEquationParabola}). The mathematical details are reported in Appendix \ref{App:Sigma2}, 
where it is proved that
\begin{equation}
\label{Eq:NumericalResults:Parabola:OffAxis.1}
\begin{array}{l}
\displaystyle
\sigma_2\,=\,3 + 3 X^2 + Y (6 + Y)\,,
\end{array}
\end{equation}
so that the third of Eq.~(\ref{Eq:Parameters.16}) leads to
\begin{equation}
\label{Eq:NumericalResults:Parabola:OffAxis.2}
\begin{array}{l}
\displaystyle
A\,=\,
X^2\,+\,(Y+1)^2\,.
\end{array}
\end{equation}
Then, the complete solution of the off-axis dressing problem turns out to be
\begin{equation}
\label{Eq:NumericalResults:Parabola:OffAxis.3}
\left\{
\begin{array}{l}
\displaystyle
\zeta(X,Y)\,=\, 2Y\,,\\
\eta(X,Y)\,=\,2X\,,\\
A(X,Y)\,=\,X^2\,+\,(Y+1)^2\,.
\end{array}
\right.
\end{equation}
It is worth checking that Eq.~(\ref{Eq:NumericalResults:Parabola:OffAxis.3}) actually satisfies Eq.~(\ref{Eq:Parameters.16}).
This can be done by substitution and on taking the following explicit expressions (obtained with the help of {\em Mathematica}):
\begin{equation}
\label{Eq:NumericalResults:Parabola:OffAxis.4}
\left\{
\begin{array}{l}
\displaystyle
\tau_4\,=\,16\, Y\, (27\, X^2 \,+ \,Y^3),\\
\tau_6\,=\,-16\, (729\, X^4\, +\, 540\, X^2 Y^3 \,-\, 8\, Y^6)\,,
\end{array}
\right.
\end{equation}
into account. 

Equation~(\ref{Eq:NumericalResults:Parabola:OffAxis.3}) is {\em the} solution of the problem. 
It might be worth seeing it at work. 
In~Fig. \ref{Fig:ParabolaSaddleExamplePhase}a, a typical scenario in which three contributive saddles are going to coalesce is shown: the observation point is $P\equiv \left(\dfrac{23}{500},\,\dfrac 15\right)$, near the right side of the caustic. In~Fig. \ref{Fig:ParabolaSaddleExamplePhase}b, the corresponding behaviour of the phase function $R^2(\varphi)$ is shown (open circles) together with the function $\dfrac{\varphi^4}4 - \zeta \dfrac{\varphi^2}2 - \eta \varphi + A$ (solid curve). As it can see, the choice of $(A,\zeta,\eta)$ given by 
Eq.~(\ref{Eq:NumericalResults:Parabola:OffAxis.3}) again provides a good fit. 
\begin{figure}
\centerline{
\begin{minipage}{3.5cm}
\centerline{\includegraphics[width=3.5cm,angle=-0]{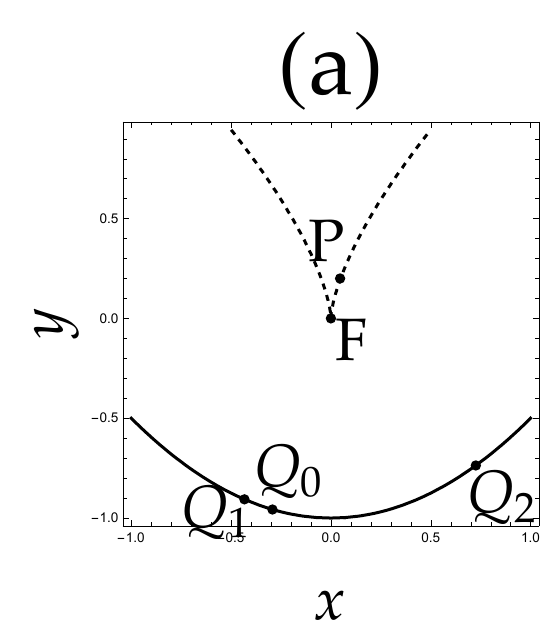}}
\end{minipage}
\hspace*{4mm}
\begin{minipage}{4.5cm}
\centerline{\includegraphics[width=4.5cm,angle=-0]{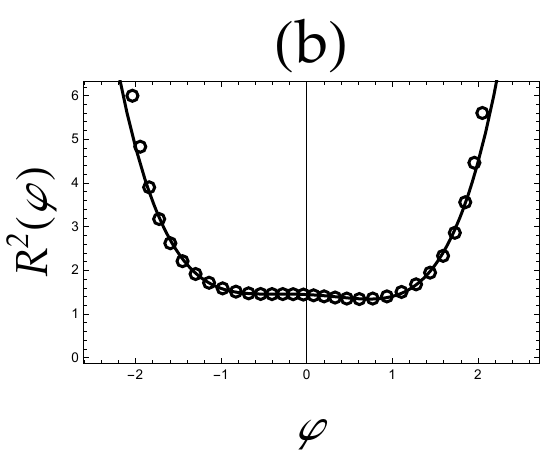}}
\end{minipage}
}
\caption{(a): a  typical example of a three coalescing saddles arrangement, with $P\equiv \left(\dfrac{23}{500},\,\dfrac 15\right)$. (b): behaviours of the corresponding phase function $R^2(\varphi)$ (open circles) and of the fourth-order polynomial approximation $\dfrac{\varphi^4}4 - \zeta \dfrac{\varphi^2}2 - \eta \varphi + A$ (solid curve), with the fit parameters $(A,\zeta,\eta)$ being given by Eq.~(\ref{Eq:NumericalResults:Parabola:OffAxis.3}).}
\label{Fig:ParabolaSaddleExamplePhase}
\end{figure}
\begin{figure}[!ht]
\centerline{\includegraphics[width=8cm,angle=-0]{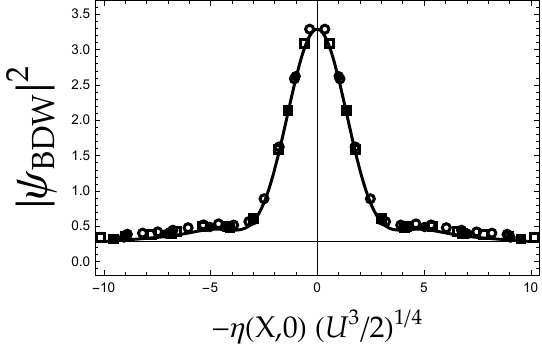}}
\caption{Behavior of the intensity $|\psi_{\rm BDW}|^2$ along the $X$-axis, obtained via  Eq.~(\ref{Eq:FresnelPropagatorConvolution.3.New.5}), for $U=500$ (open circles), $U=1000$
(open squares)  $U=2000$ (black circles) and $U=3000$ (black squares).
Intensity values  are plotted against the normalized variable $s_1= -\eta(X,Y)\,\left({U^3}/2\right)^{1/4}$,
where $\eta(X,0)=2X$, according to Eq.~(\ref{Eq:NumericalResults:Parabola:OffAxis.3}). 
Solid curve: function $|\mathrm{Pe}(s_1,0)|^2$.}
\label{Fig:PearceyScalingLawBis}
\end{figure}
In the last two figures we are going to present, the scaling law given into 
Eq.~(\ref{Eq:3SaddleContribution}) will be tested in the most general case. In particular, in~Fig. \ref{Fig:PearceyScalingLawBis} 
the values of $|\psi_{\rm BDW}|^2$ have now been numerically computed along the $X$ axis and, again after a normalization by 
the factor $\pi^2\sqrt{2U}$, are plotted against the scaled variable 
$s_1=-\eta(X,0)\,\left(\dfrac {U^3}2\right)^{1/4}$, for the same values of the Fresnel number chosen for 
Fig. \ref{Fig:PearceyScalingLaw}, i.e., $U=500,\,1000,\,2000,\,3000$. 
The solid curve now represents the function $|\mathrm{Pe}(s_1,0)|^2$.
Equation~(\ref{Eq:NumericalResults:Parabola:OffAxis.3})
will now be employed to validate the most general form of the asymptotics in Eq.~(\ref{Eq:3SaddleContribution}).
To this end, readers must be adviced that a further, nontrivial computational problem is represented by the numerical evaluation of the Pearcey function for typical pairs $(s_1,s_2)\in\mathbb{R}^2$. Such a practical problem has been tackled and solved in several ways since the original Pearcey paper~\cite{Pearcey/1946}. Here, the algorithm developed in~\cite{Borghi/2007,Borghi/2016b} has been employed to generate the high density 2D map of the function $|{\rm Pe}(s_1,s_2)|^2$ in~Fig.~\ref{Fig:PearceyScalingLawTer} (yellow surface).
\begin{figure}[!ht]
\centerline{\includegraphics[width=8cm,angle=-0]{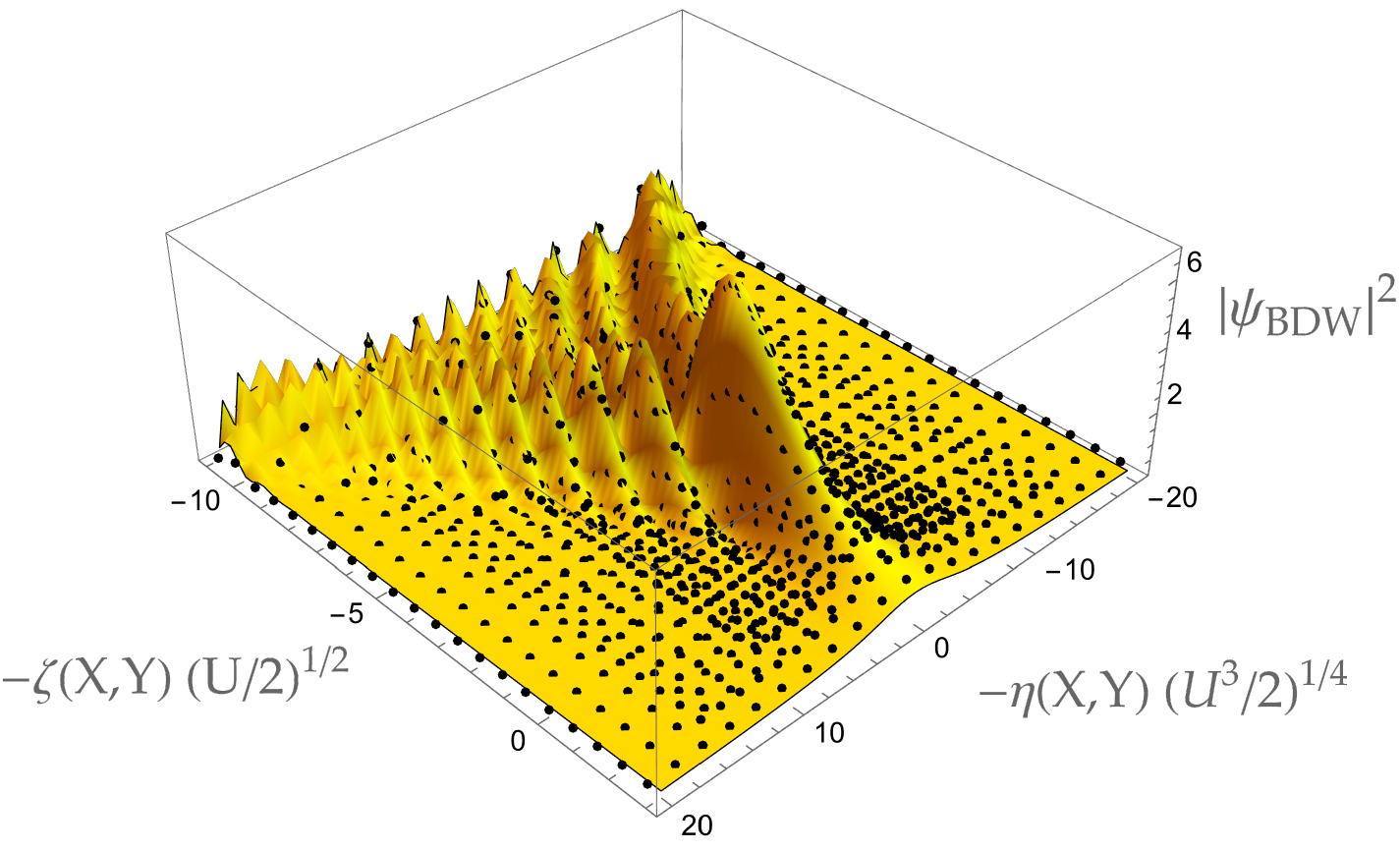}}
\caption{Behavior of the intensity $|\psi_{\rm BDW}|^2$ across the $(X,Y)$-plane, obtained via  Eq.~(\ref{Eq:FresnelPropagatorConvolution.3.New.5}). Dots represent intensity values evaluated for $U=500$, $U=1000$, and $U=1500$, which are plotted against the normalized variables $s_1= -\eta(X,Y)\,\left({U^3}/2\right)^{1/4}$ and $s_2= -\zeta(X,Y)\,\left(U/2\right)^{1/2}$,
with $\eta(X,Y)$ and $\zeta(X,Y)$ being given by Eq.~(\ref{Eq:NumericalResults:Parabola:OffAxis.3}). 
The yellow  surface represents the function $|\mathrm{Pe}(s_1,s_2)|^2$.}
\label{Fig:PearceyScalingLawTer}
\end{figure}

In the same figure, 
dots represent the values of $|\psi_{\rm BDW}|^2$ computed for $U=500$, $U=1000$, and $U=1500$. These values are  plotted against the normalized variables $s_1= -\eta(X,Y)\,\left({{U^3}/2}\right)^{1/4}$ and $s_2= -\zeta(X,Y)\,\left({U/2}\right)^{1/4}$, where again
$\eta(X,Y)$ and $\zeta(X,Y)$ are given by Eq.~(\ref{Eq:NumericalResults:Parabola:OffAxis.3}).

The dress fits like a glove.

\section{Conclusions}
\label{Sec:Conclusions}

The description of light diffraction phenomena from the  optics perspective requires solving a fundamental problem: to find the real arguments that feed all involved diffraction catastrophes, starting from the singularity arrangements and thereby allowing wave optics to be ``added'' to geometric optics.  

For the lowest order catastrophe, the fold, a complete solution to this nontrivial problem has been provided in the past.
Cusp represents the next step in the hierarchy of stable singularities. Unlike folds, the technical problems that arise when decorating cusp singularities with the corresponding diffraction catastrophe (Pearcey function) are much more difficult. To our knowledge, this problem has not been explicitly solved, except for certain highly symmetric scenarios.

In the present paper, the recently developed paraxial BDW's theory~\cite{Borghi/2015,Borghi/2016} has been used to design an algorithm aiming at the complete solution of the ``cusp dressing problem'', i.e., the complete determination of the uniform asymptotics of the phase integrals with {three} coalescing saddles.
In particular, Newton's identities for quadratic and cubic equations were used to prove that
the real arguments of the involved Pearcey function must satisfy a nonlinear algebraic system whose exact 
solution has also been found explicitly.

A single but significant numerical experiment has been carried out to check the validity and effectiveness of the proposed algorithm:
the complete catastrophe optics characterization of the field generated at observation points near the cusp singularity, when plane waves are diffracted by a sharp-edged parabolic profile. 
2D and 3D visual checks of the goodness of the uniform asymptotics derived in this way have also been provided. 

It must be noted how the applicability of the algorithm presented here be not limited to sharp-edge diffraction, because the solution of the dressing problem requires only the values of the phase function assumed at the coalescing saddle points. 
Accordingly, we hope what we have found here could also be applied to more general scenarios of natural focusing of light, where CO is the most mathematically suitable language.
Under such a perspective, the extension of our algorithm to higher-order cuspoid diffraction catastrophes, such as swallowtail and butterfly~\cite{Nye/1999}, would also be highly desirable. 
A rather challenging task, which is worthy of being tackled in the future. 

\appendix


\section{Proof of Eq.~(\ref{Eq:Parameters.15})}
\label{App:NonlinearSystem}

Consider the square of both sides of Eq.~(\ref{Eq:Parameters.14}), i.e.,
\begin{equation}
\label{Eq:Parameters.14Squared}
\begin{array}{l}
\displaystyle
R_k^4\,=\,A^2 - 2 A \eta \xi_k - 
( A \zeta - \eta^2) \xi_k^2 \\
\\ + \zeta \eta \xi_k^3 + 
 \dfrac{2A +\zeta^2}2\,\xi_k^4 + 
 \dfrac{\eta \xi_k^5}2 - 
 \dfrac{\zeta \xi_k^6}4 + 
 \dfrac{\xi_k^8}{16}\,.
\end{array}
\end{equation}
On summing side by side Eq. (\ref{Eq:Parameters.14Squared}) over $k=0,1,2$, after rearranging we have
\begin{equation}
\label{Eq:Parameters.14SquaredBis}
\begin{array}{l}
\displaystyle
\sigma_4\,=\,3A^2\,-\,A{\zeta^2}\,+\,\dfrac94\,\zeta\eta^2\,+\,\dfrac{\zeta^4}8\,,
\end{array}
\end{equation}
where use has been made of the following Newton identities applied to Eq.~(\ref{Eq:Parameters.13}):
\begin{equation}
\label{Eq:Parameters.NewtonTheor}
\left\{
\begin{array}{l}
\displaystyle
\sum_k \xi_k =0\,,\\
\displaystyle
\sum_k\xi_k^2=2\zeta\,,\\ 
\displaystyle
\sum_k \xi_k^3=3\eta\,,\\
\displaystyle
\sum_k\xi_k^4=2\zeta^2\,,\\
\displaystyle
\sum_k\xi_k^5=5\zeta\eta\,,\\ 
\displaystyle
\sum_k\xi_k^6=2\zeta^3+3\eta^2\,, \\
\displaystyle
\sum_k\xi_k^8=\,2(\zeta^4+4\zeta\eta^2)\,.
\end{array}
\right.
\end{equation}
The third power of of both sides of Eq.~(\ref{Eq:Parameters.14}) then gives
\begin{equation}
\label{Eq:Parameters.14Cubic}
\begin{array}{l}
\displaystyle
R_k^6\,=\,
A^3 - 
3 A^2 \eta \xi_k +
3 A \left(\eta^2 -\dfrac{A\zeta}2\right)\,\xi_k^2 + 
( 3 A \zeta - \eta^2)\eta \xi_k^3 
\\+ 
\dfrac 34 \left(A^2 + A \zeta^2 - 2 \zeta \eta^2\right)\, \xi_k^4
 - \dfrac 34 \left(\zeta^2+2A\right) \eta \xi_k^5 \\
- \dfrac 18 \left(6A\zeta+\zeta^3-6\eta^2\right) \xi_k^6 
+ \dfrac  34 \zeta \eta \xi_k^7 
+\dfrac 3{16} (A  + \zeta^2) \xi_k^8 \\
- \dfrac 3{16} \eta \xi_k^9 - \dfrac 3{32}  \zeta \xi_k^{10} 
+ \dfrac{\xi_k^{12}}{64},
\qquad k\,=\,0,1,2\,.
\end{array}
\end{equation}
Finally, on proceeding as before, after taking Eq.~(\ref{Eq:Parameters.NewtonTheor}) into account, together with
\begin{equation}
\label{Eq:Parameters.NewtonTheorBis}
\left\{
\begin{array}{l}
\displaystyle
\sum_k\xi_k^7=\,7 \zeta^2 \eta\,,\\
\displaystyle
\sum_k\xi_k^9=\,3 (3 \zeta^3 \eta + \eta^3)\,,\\
\displaystyle
\sum_k\xi_k^{10}=\,2 \zeta^5 + 15 \zeta^2 \eta^2\,,\\
\displaystyle
\sum_k\xi_k^{12}=\,2 \zeta^6 + 24 \zeta^3 \eta^2 + 3 \eta^4\,,
\end{array}
\right.
\end{equation}
long but straightforward algebra eventually leads to
\begin{equation}
\label{Eq:Parameters.14CubicBis}
\begin{array}{l}
\displaystyle
\sigma_6\,=\, 3 A^3\,-\,\dfrac 32 A^2\zeta^2\,+\,\dfrac 38 A\zeta^4\,-\,\dfrac{\zeta^6}{32}\,+\,\dfrac{27}{4} A\zeta\eta^2\,\\
-\,\dfrac{51}{32}\zeta^3\eta^2\,-\,\dfrac{81}{64} \eta^4\,.
\end{array}
\end{equation}
Equations (\ref{Eq:Parameters.14SquaredBis}) and (\ref{Eq:Parameters.14CubicBis}) are the second and the third row, respectively, of the system in Eq. (\ref{Eq:Parameters.15}).

\section{Proof of Eq.~(\ref{Eq:Parameters.16})}
\label{App:SolvingSystem}

For simplicity, system~(\ref{Eq:Parameters.15}) is rewritten below,
\begin{equation}
\label{App:SolvingSystem.1}
\left\{
\begin{array}{l}
\displaystyle
\sigma_2\,=\,3A\,-\,\dfrac{\zeta^2}2\,,\\
\\
\sigma_4\,=\,3A^2\,-\,A{\zeta^2}\,+\,\dfrac94\,\zeta\eta^2\,+\,\dfrac{\zeta^4}8\,,\\
\\
\sigma_6\,=\, 3 A^3\,-\,\dfrac 32 A^2\zeta^2\,+\,\dfrac 38 A\zeta^4\,-\,\dfrac{\zeta^6}{32}\,+\,\dfrac{27}{4} A\zeta\eta^2\\
\,-\,\dfrac{51}{32}\zeta^3\eta^2\,-\,\dfrac{81}{64} \eta^4\,.
\end{array}
\right.
\end{equation}
First of all, $A$ can be eliminated by recasting the first row as follows:
\begin{equation}
\label{App:SolvingSystem.2}
\begin{array}{l}
\displaystyle
A\,=\,\dfrac{\sigma_2}3\,+\,\dfrac{\zeta^2}6\,,
\end{array}
\end{equation}
so that, after substitution, the remaining equations become
\begin{equation}
\label{App:SolvingSystem.3.1}
\begin{array}{l}
\displaystyle
\zeta^4\,+54\,\zeta\,\eta^2\,=\,24\sigma_4-8\sigma_2^2\,,
\end{array}
\end{equation}
and
\begin{equation}
\label{App:SolvingSystem.3.2}
\begin{array}{l}
\displaystyle
2\zeta^6\,-\,270\,\zeta^3\,\eta^2\,+\,24\,\sigma_2\,(\zeta^4\,+54\,\zeta\,\eta^2)\,+64\,\sigma^3_2-729\,\eta^4\,=\,\\
\,=\,576\,\sigma_6\,,
\end{array}
\end{equation}
respectively. A further substitution from Eq.~(\ref{App:SolvingSystem.3.1}) into Eq.~(\ref{App:SolvingSystem.3.2}), after rearranging gives
\begin{equation}
\label{App:SolvingSystem.3.3}
\begin{array}{l}
\displaystyle
2\zeta^6\,-\,270\,\zeta^3\,\eta^2\,-\,729\,\eta^4\,=\,576\,(\sigma_6\,-\,\sigma_4\sigma_2)\,+\,128\,\sigma^3_2\,,
\end{array}
\end{equation}
in such a way that the system for the pair $(\zeta,\eta)$ takes on the form
\begin{equation}
\label{App:SolvingSystem.4}
\left\{
\begin{array}{l}
\displaystyle
\zeta^4\,+54\,\zeta\,\eta^2\,=\,\tau_4\,,
\\
2\zeta^6\,-\,270\,\zeta^3\,\eta^2\,-\,729\,\eta^4\,=\,\tau_6\,,
\end{array}
\right.
\end{equation}
where the auxiliary quantities
\begin{equation}
\label{App:SolvingSystem.5}
\left\{
\begin{array}{l}
\displaystyle
\tau_4\,=\,24\sigma_4-8\sigma_2^2\,\\
\tau_6\,=\,576\,(\sigma_6\,-\,\sigma_4\sigma_2)\,+\,128\,\sigma^3_2\,,
\end{array}
\right.
\end{equation}
have been introduced. 
System~(\ref{App:SolvingSystem.4}) can further be reduced. From the first equation we have
\begin{equation}
\label{App:SolvingSystem.5}
\begin{array}{l}
\displaystyle
\eta^2\,=\,\dfrac{\tau_4\,-\,\zeta^4}{54\,\zeta}\,,
\end{array}
\end{equation}
which, once substituted into the second equation, after rearranging gives the following single algebraic equation for $\zeta$:
\begin{equation}
\label{App:SolvingSystem.6}
\begin{array}{l}
\displaystyle
27\,\zeta^8\,-\,18\tau_4\zeta^4-4\tau_6\zeta^2-\tau^2_4=0\,.
\end{array}
\end{equation}
Equations (\ref{App:SolvingSystem.6}), (\ref{App:SolvingSystem.5}), and (\ref{App:SolvingSystem.2})
constitute the system in Eq. (\ref{Eq:Parameters.16}).

\section{Proof of Eq.~(\ref{Eq:NumericalResults:Parabola:OffAxis.2})}
\label{App:Sigma2}

On denoting $t_k$ the $k$th solution of Eq. (\ref{Eq:SaddleEquationParabola}), with $k=0,1,2$, the corresponding quantity $R^2_k$ turns out to be
\begin{equation}
\label{App:Sigma2.1}
\begin{array}{l}
\displaystyle
R^2_k\,=\,\overline{PQ_k}^2\,=\,(t_k-X)^2\,+\,\left(\dfrac{t^2_k}2-Y-1\right)^2\,=\,\\
\,=\,X^2\,+\,(Y+1)^2\,+\,\dfrac{t^4_k}4\,-2Y\,t^2_k\,-2X\,t_k\,,
\end{array}
\end{equation}
so that
\begin{equation}
\label{App:Sigma2.2}
\begin{array}{l}
\displaystyle
\sigma_2\,=\,\sum_k\,R^2_k\,=\,3X^2\,+\,3(Y+1)^2\,+\,\dfrac 14{\sum_kt^4_k}\,\\
\\
\displaystyle
-2Y{\sum_kt^2_k}\,-2X{\sum_kt_k}\,.
\end{array}
\end{equation}
Finally, on again using Newton's identities into Eq. (\ref{Eq:Parameters.NewtonTheor}), but now applied to 
Eq. (\ref{Eq:SaddleEquationParabola}), trivial algebra leads to Eq.~(\ref{Eq:NumericalResults:Parabola:OffAxis.2}). 

\acknowledgements
I wish to thank Gabriella Cincotti and Turi Maria Spinozzi for their invaluable help during the preparation of the manuscript.

%

%

\end{document}